\documentclass[epj]{svjour}
\usepackage{color}
\definecolor{darkgreen}{rgb}{0,0.5,0}
\definecolor{purple}{rgb}{0.5,0,0.5}
\definecolor{nblue}{rgb}{0.0,0.0,0.50}
\definecolor{scarlet}{rgb}{1.0,0.2,0}
\usepackage[colorlinks=true,
            pdfstartview=FitV,
        linkcolor=purple,
        citecolor=purple,
        urlcolor=blue,
        pdftitle={Lattice Variational Approach},
        pdfpagemode=None
       ]{hyperref}
\usepackage{graphicx}
\usepackage{graphics}
\renewcommand{\thesection}{\arabic{section}.}
\newcommand{\sfrac}[2]{\mbox{\footnotesize $\displaystyle \frac{#1}{#2}$}} 
\newcommand{\nlsim}{\mathrel{\rlap{\lower4pt\hbox{\hskip0pt$\sim$}} 
 \raise1pt\hbox{$<$}}}           
\newcommand{\ngsim}{\mathrel{\rlap{\lower4pt\hbox{\hskip0pt$\sim$}} 
 \raise1pt\hbox{$>$}}}           
\begin{document}
\title{Mind the gap\thanks{CDR is grateful to the organisers for their hospitality and acknowledges fruitful conversations with O.~Lakhina and P.\,C.~Tandy.  This work was supported by: US Department of Energy, Office of Nuclear Physics, contract nos.\ DE-AC02-06CH11357 and DE-FG02-00ER41135; Austrian Science Fund FWF, Schr\"odinger-R\"uckkehrstipendium R50-N08; and benefited from the facilities of the ANL Computing Resource Center and Pittsburgh's NSF Terascale Computing System.}
}
\author{M.\,S.~Bhagwat\inst{1} \and A.~Krassnigg\inst{2} \and P.~Maris\inst{3} \and C.\,D.~Roberts\inst{1}
}                     
%
%
\institute{Physics Division, Argonne National Laboratory,
Argonne, IL, 60439, USA \and Fachbereich Theoretische Physik, Universit\"at Graz, A-8010 Graz, Austria \and Department of Physics and Astronomy, University of Pittsburgh, PA 15260, USA}
%
\date{\today}
%
\abstract{
In this summary of the application of Dyson-Schwinger equations to the theory and phenomenology of hadrons, some deductions following from a nonperturbative, symmetry-preserving truncation are highlighted, notable amongst which are results for pseudoscalar mesons.  We also describe inferences from the gap equation relating to the radius of convergence of a chiral expansion, applications to heavy-light and heavy-heavy mesons, and quantitative estimates of the contribution of quark orbital angular momentum in pseudoscalar mesons; and recapitulate upon studies of nucleon electromagnetic form factors.
\PACS{
    {24.85.+p}{Quarks, gluons, and QCD in nuclei and nuclear processes} \and 
    {12.38.Lg}{Other nonperturbative calculations}
     } 
} 
\maketitle
\section{Introduction}
\label{intro}
The world's experimental hadron physics facilities are providing data of unprecedented accuracy with an enormous potential to impact on our understanding of the basic features of the strong interaction; e.g., \cite{gao03,Burkert:2004sk,Beise:2004py,arrington}.  The explanation of that data in terms of QCD's key elements is an important task for contemporary theory.  It is likewise vital for theory to link current observations to a future path of discovery.  Success with that requires flexible tools, which can rapidly provide an intuitive understanding of information in hand and simultaneously anticipate its likely consequences.  Models, parametrisations and truncations of QCD play this role.  Such theory can also identify and explore novel possibilities within hadron physics, whose experimental verification could test the foundations of QCD.  

Prominent among such tools are QCD's Dyson-Schwin\-ger equations (DSEs), truncations thereof, and models based on this complex of integral equations.  Glancing back at the last proceedings volume produced for this series of conferences \cite{Maris:2002mt}, it is apparent that considerable progress has been made in the interim.  We will provide a snapshot of that herein, which is augmented by other contributions to this volume.
 
\section{Gap equation}
\label{sec:1}
In the presence of dynamical chiral symmetry breaking (DCSB); i.e., the generation of mass \textit{from nothing}, a study of hadrons must begin with QCD's gap equation:
\begin{eqnarray}
S(p)^{-1} & =&  Z_2 \,(i\gamma\cdot p + m^{\rm bm}) + \Sigma(p)\,, \label{gendse} \\
\Sigma(p) & = & Z_1 \int^\Lambda_q\! g^2 D_{\mu\nu}(p-q) \frac{\lambda^a}{2}\gamma_\mu S(q) \Gamma^a_\nu(q,p) , \label{gensigma}
\end{eqnarray}
where $\int^\Lambda_q$ represents a Poincar\'e invariant regularisation of the integral, with $\Lambda$ the regularisation mass-scale \cite{mrt98}, $D_{\mu\nu}(k)$ is the dressed-gluon propagator, $\Gamma_\nu(q,p)$ is the dressed-quark-gluon vertex, and $m^{\rm bm}$ is the quark's $\Lambda$-dependent bare current-mass.  The vertex and quark wave-function renormalisation constants, $Z_{1,2}(\zeta^2,\Lambda^2)$, depend on the gauge parameter.  

The solution of the gap equation can be written 
\begin{eqnarray} 
 S(p) & =&  
Z(p^2,\zeta^2)/[i\gamma\cdot p + M(p^2)]
%
\label{Sgeneral}
\end{eqnarray} 
wherein the mass function, $M(p^2)$, is momentum-dependent but independent of the renormalisation point.  The solution is obtained from (\ref{gendse}) augmented by the renormalisation condition
\begin{equation}
\label{renormS} \left.S(p)^{-1}\right|_{p^2=\zeta^2} = i\gamma\cdot p +
m(\zeta^2)\,,
\end{equation}
where $m(\zeta^2)$ is the renormalised (running) mass: 
\begin{equation}
Z_2(\zeta^2,\Lambda^2) \, m^{\rm bm}(\Lambda) = Z_4(\zeta^2,\Lambda^2) \, m(\zeta^2)\,,
\end{equation}
with $Z_4$ the Lagrangian-mass renormalisation constant.  In QCD the chiral limit is strictly defined by \cite{mrt98}
\begin{equation}
\label{limchiral}
Z_2(\zeta^2,\Lambda^2) \, m^{\rm bm}(\Lambda) \equiv 0 \,, \forall \Lambda^2 \gg \zeta^2 \,,
\end{equation}
which states that the renormalisation-point-invariant cur\-rent-quark mass $\hat m = 0$.

In connection with (\ref{gendse}) \& (\ref{gensigma}) it has recently been established \cite{Chang:2006bm} that on a bounded, measurable domain of non-negative current-quark mass, realistic models for the kernel can simultaneously admit two inequivalent DCSB solutions and a solution that is unambiguously connected with the realisation of chiral symmetry in the Wigner mode.  The Wigner solution and one of the DCSB solutions are destabilised by a current-quark mass and both disappear when that mass exceeds a critical value.  This critical value also bounds the domain on which the surviving DCSB solution possesses a chiral expansion and can therefore be viewed as an upper limit on the domain within which a perturbative expansion in the current-quark mass around the chiral limit is uniformly valid for physical quantities.  This critical mass is typically $\hat m_{\rm cr} \sim 60$--$70\,$MeV, which for a flavour-nonsinglet $0^{-}$ meson constituted of equal mass quarks corresponds to a mass $m_{0^{-}} \sim 0.45\,$GeV \cite{andreaspi}.  In arguing this case, properties of the two DCSB solutions of the gap equation that enable a valid definition of $\langle \bar q q \rangle$ in the presence of a nonzero current-mass were employed.  The behaviour of this condensate indicates that the essentially dynamical component of chiral symmetry breaking decreases with increasing current-quark mass, following the trend predicted by the constituent-quark $\sigma$-term (\cite{HUGS05}, Sec.\,5.2.2).

\section{Symmetry preserving truncation}
One is forced to begin with (\ref{gendse}) \& (\ref{gensigma}) by the axial-vector Ward-Takahashi identity; viz., 
\begin{eqnarray}
\nonumber P_\mu \Gamma_{5\mu}(k;P)  &= &S(k_+)^{-1} i \gamma_5
+  i \gamma_5 S(k_-)^{-1}\\
&& - 2 i m(\zeta^2) \Gamma_5(k;P)\,,
\label{avwtim}
\end{eqnarray}
written here for a quark and antiquark of equal current-mass, wherein the axial vector vertex is determined by the inhomogeneous Bethe-Salpeter equation
\begin{equation}
\label{avbse}
\left[\Gamma_{5\mu}(k;P)\right]_{tu}
 =  Z_2 \left[\gamma_5\gamma_\mu\right]_{tu} + \int^\Lambda_q [\chi(q;P)]_{sr}
 K_{tu}^{rs}(q,k;P),
\end{equation}
with $\chi(k;P)=S(k_+) \Gamma_{5\mu}(k;P) S(k_-)$, $k_\pm = k \pm P/2$ and the colour- and Dirac-matrix structure of the elements in the equation is denoted by the indices $r,s,t,u$.  In (\ref{avbse}), $K(q,k;P)$ is the fully-amputated quark-antiquark scattering kernel.  If one knows the form of $K$ then the nature of the interaction between quarks in QCD is completely understood.  The pseudoscalar vertex in (\ref{avwtim}) is prescribed by an analogue of (\ref{avbse}).

Every pseudoscalar meson appears as a pole contribution to the axial-vector and pseudoscalar vertices \cite{mrt98}; viz., 
\begin{eqnarray}
\nonumber && \Gamma_{5 \mu}(k;P)\stackrel{{P^2 \simeq -m_{\pi_n}^2 }}{=}   \frac{f_{\pi_n} \, P_\mu}{P^2 + 
m_{\pi_n}^2} \Gamma_{\pi_n}(k;P) 
+ \; \Gamma_{5 \mu}^{{\rm reg}}(k;P) \,, \\ 
\label{genavv} \\
\nonumber && i\Gamma_{5 }(k;P)\stackrel{P^2 \simeq -m_{\pi_n}^2}{=}  
 \frac{\rho_{\pi_n}(\zeta^2) }{P^2 + 
m_{\pi_n}^2} \Gamma_{\pi_n}(k;P)
 + \; i\Gamma_{5 }^{{\rm reg}}(k;P) \,, \\ \label{genpv} 
\end{eqnarray}
where the $\Gamma^{{\rm reg}}$ are regular in the neighbourhood of this pole, $\Gamma_{\pi_n}(k;P)$ represents the bound state's canonically normalised Bethe-Salpeter amplitude: 
\begin{eqnarray} 
\nonumber
\lefteqn{
\Gamma_{\pi_n}(k;P) =  \gamma_5 \left[ i E_{\pi_n}(k;P) + \gamma\cdot P F_{\pi_n}(k;P) \right. }\\
&+& \left.
    \gamma\cdot k \,k \cdot P\, G_{\pi_n}(k;P) + 
\sigma_{\mu\nu}\,k_\mu P_\nu \,H_{\pi_n}(k;P)  \right] \! , \label{genpibsa} \end{eqnarray}
and
\begin{eqnarray} 
\label{fpin} f_{\pi_n}   P_\mu &=& Z_2\,{\rm tr} \int^\Lambda_q 
 \gamma_5\gamma_\mu\, \chi_{\pi_n}(q;P) \,, \\
\label{cpres} i  \rho_{\pi_n}\!(\zeta^2)\,   &=& Z_4\,{\rm tr} 
\int^\Lambda_q \gamma_5 \, \chi_{\pi_n}(q;P)\,,
\end{eqnarray} 
wherein the Bethe-Salpeter wave function is
\begin{eqnarray}
\lefteqn{\chi_{\pi_n}(k;P) = S(k_+) \Gamma_{\pi_n}(k;P) S(k_-)}\\
& = & 
\nonumber
\lefteqn{
 \gamma_5 \left[ i {\cal E}_{\pi_n}(k;P) + \gamma\cdot P {\cal F}_{\pi_n}(k;P) \right. }\\
&+& \left.
    \gamma\cdot k \,k \cdot P\, {\cal G}_{\pi_n}(k;P) + 
\sigma_{\mu\nu}\,k_\mu P_\nu \,{\cal H}_{\pi_n}(k;P)  \right] \! . \label{genpibswf} 
\end{eqnarray}
In (\ref{genavv}) -- (\ref{genpibswf}), $\pi_0$ denotes the lowest-mass pseudo\-sca\-lar and increasing $n$ labels bound-states of increasing mass.

Equation (\ref{avwtim}) is a statement of chiral symmetry and the manner by which it is broken, explicitly and dynamically.  It relates the solution of the two-body problem, (\ref{avbse}), in quantum field theory to the one-body problem, (\ref{gendse}).  QCD is violated in any approach that does not preserve (\ref{avwtim}).  A weak-coupling expansion guarantees (\ref{avwtim}).  However, one cannot study bound-states in perturbation theory, nor QCD's fundamental emergent phenomena; viz., confinement and DCSB.  Fortunately, as related in \cite{Maris:2002mt}, there is at least one symmetry-preserving truncation of the DSEs that is nonperturbative in the coupling \cite{munczek,truncscheme,detmoldvertex,Bhagwat:2004hn}.  This fact has enabled the proof of exact results in QCD and importantly in addition their illustration using practical truncations to which the corrections are quantifiable.

An exemplar: for a flavour-nonsinglet pseudoscalar meson\footnote{For notational simplicity we've written this identity for mesons constituted from a quark and antiquark with the same current-mass.  The unequal mass case is little different.} \cite{mrt98,Holl:2004fr}
\begin{equation}
\label{gengmor}
f_{\pi_n} \, m_{\pi_n}^2 = 
2 m(\zeta^2)\,\rho_{\pi_n}(\zeta^2) \,.
\end{equation}
The Gell-Mann--Oakes--Renner relation is a corollary of (\ref{gengmor}).  Moreover, in deriving this expression, no assumptions are made about the current-quark mass of the  constituents.  Hence an analogue is equally valid for a meson containing one heavy-quark, a heavy-light system, and also heavy-heavy mesons.  

Furthermore, the validity of (\ref{gengmor}) is not restricted to the ground state.  This entails \cite{Holl:2004fr} that in the chiral limit
\begin{equation}
\label{fpinzero}
f_{\pi_n}^0 \equiv 0\,, \forall \, n\geq 1\,;
\end{equation}
viz., Goldstone modes are the only pseudoscalar mesons to possess a nonzero leptonic decay constant in the chiral limit when chiral symmetry is dynamically broken.  The decay constants of all other pseudoscalar mesons on this trajectory, e.g., radial excitations, vanish.  This result is consistent with model studies; e.g., \cite{dominguez,LeYaouanc:1984dr,Llanes-Estrada:1999uh,volkov,Lucha:2006rq}.  On the flip side, in the absence of DCSB, the leptonic decay constant of \emph{each} such pseudoscalar meson vanishes in the chiral limit; namely, (\ref{fpinzero}) is true $\forall n\geq 0$.

From the perspective of quantum mechanics, (\ref{fpinzero}) is a surprising fact.  The leptonic decay constant for $S$-wave states is typically proportional to the wave func\-tion at the origin.  Compared with the ground state, this is smaller for an excited state because the wave function is broader in configuration space and wave func\-tions are normalised.  However, it is a modest effect; e.g., consider the $e^+ e^-$ decay of vector mesons, for which a calculation in relativistic quantum mechanics based on light-front dynamics \cite{deMelo:2005cy} yields $f_{\rho_1}/f_{\rho_0}\approx 0.6$, consistent with the value inferred from experiment.  [The expression for $f_V$ is analogous to (\ref{fpin}), see \cite{Ivanov:1998ms}.]  As apparent in a recent lattice simulation \cite{McNeile:2006qy}, (\ref{fpinzero}) can be realised in a framework if, and only if, (\ref{avwtim}) holds true and is veraciously realised.

The two-photon decay of $0^{-+}$ mesons provides another important example.  In the presence of DCSB the ground state neutral pseudoscalar meson decays predominantly into two photons.  So long as one employs a nonperturbative truncation that preserves chiral symmetry and the pattern of its dynamical breaking, the ground state's two photon decay is described in the chiral limit by a coupling \cite{Bando:1993qy,Roberts:1994hh,Maris:1998hc,Maris:2002mz}
\begin{equation}
\label{anomalycouple}
g_{\pi_0^0 \gamma\gamma} := {\cal T}_{\pi_n^0}(-m_{\pi_n}^2=0,Q^2=0) = \frac{1}{2} \frac{1}{f_{\pi_0}}\,,
\end{equation}
where ${\cal T}$ is the scalar function appearing in the matrix element.  Equation\,(\ref{anomalycouple}) is the most widely known consequence of the Abelian anomaly.  

Now, given that $f_{\pi_n \neq 0} \equiv 0$ in the chiral limit, it is natural to ask whether the $\pi_{n\neq 0} \to \gamma\gamma$ transition is affected.  Since rainbow-ladder is the leading order in a symmetry preserving truncation, it was used in \cite{Holl:2005vu} to provide a model-independent analysis of this process for arbitrary $n$.  Therein, amongst other things, (\ref{anomalycouple}) is generalised and it is proved that
\begin{eqnarray}
\nonumber \lefteqn{
{\cal T}_{\pi_n^0}(-m_{\pi_n}^2,Q^2)}\\
&&  \stackrel{Q^2\gg \Lambda_{\rm QCD}^2}{=} \frac{4\pi^2}{3}
\left[ \frac{f_{\pi_n}}{Q^2} + F_n^{(2)}(-m_{\pi_n}^2) 
\frac{\ln^{\gamma} Q^2/\omega_{\pi_n}^2}{Q^4}
\right] \!,
\label{UVnot0}
\end{eqnarray}
where: $\gamma$ is an anomalous dimension; $\omega_{\pi_n}$ is a mass-scale that gauges the momentum space width of the pseudoscalar meson; and $F_n^{(2)}(-m_{\pi_n}^2)$ is a structure-dependent constant, similar but unrelated to $f_{\pi_n}$.\footnote{With the interaction described in \protect\cite{Holl:2005vu}, $F_1^{(2)}(-m_{\pi_n}^2) \simeq -\langle\bar q q\rangle^0$, and it is generally nonzero in the chiral limit.}   It is now plain that $\forall\, n\geq 1$ 
\begin{eqnarray}
\nonumber
\lefteqn{\lim_{\hat m\to 0} \check{T}_{\pi_n^0}(-m_{\pi_n}^2,Q^2) }\\
&& \stackrel{Q^2\gg \Lambda_{\rm QCD}^2}{=} \frac{4\pi^2}{3}\left. F^{(2)}_{n }(-m_{\pi_n}^2)\frac{\ln^{\gamma} Q^2/\omega_{\pi_n}^2}{Q^4}\right|_{\hat m=0} ;
\label{UVchiralnot0}
\end{eqnarray}
namely, in the chiral limit the leading-order power-law in the transition form factor for excited state pseudoscalar mesons is O$(1/Q^4)$.  

\section{Heavy quarks}
The implications of (\ref{gengmor}) for heavy-light mesons are treated in \cite{Ivanov:1998ms,Ivanov:1997yg,Ivanov:1997iu}.  Therefore herein we focus on heavy-heavy mesons, for which (\ref{gengmor}) is equally valid.  For quark flavour $Q$, a constituent-quark spectrum-mass is defined via
\begin{equation}
M^S_Q = M_Q(p^2=\zeta^2_{\rho_0})\,,\; \zeta^2_{\rho_0}=-\sfrac{1}{4}M_{\rho_0}^2
\end{equation}
where $M_Q(p^2)$ is the renormalisation-point-invariant dres\-sed-quark mass function in (\ref{Sgeneral}) obtained as the solution of (\ref{gendse}) \& (\ref{gensigma}) with the $Q$-quark current-mass in (\ref{renormS}).\footnote{In solving the Bethe-Salpeter equation, an amplitude peaked at zero relative four-momentum weights this value of $M_Q(p^2)$ most heavily in the dressed-quark propagator.  In calculations based on \protect\cite{Maris:2002mt}, $M^S_c=2.00\,$GeV \& $M^S_b=5.34\,$GeV.}  As $\hat m_Q$ is increased, 
$M^S_Q$ becomes equivalent to the so-called pole-mass in the effective field theory for quarkonium systems; i.e., non-relativistic-QCD (NRQCD).  Now, with 
\begin{equation}
m_{\pi_{n}}^{\bar Q Q} = 2 M^S_Q \,[ 1 + \varepsilon_{\pi_n}^{\bar QQ}/M^S_Q]\,,
\end{equation}
where $\varepsilon_{\pi_n}^{\bar QQ}$ is a binding-energy that does not grow with $M^S_Q$, and using the renormalisation-group-invariance of $m(\zeta^2) \rho_{\pi_{n}}^{\bar Q Q}(\zeta^2)$,  (\ref{gengmor}) predicts
\begin{equation}
\label{QQidentity}
\rho_{\pi_{n}}^{\bar Q Q}(\zeta^2_{\rho_0}) \stackrel{\hat m_Q \to \infty}{=} f_{\pi_{n}}^{\bar Q Q} m_{\pi_{n}}^{\bar Q Q} \,.
\end{equation}
This is an identity between the pseudoscalar and pseudovector projections of the meson's Bethe-Salpeter wave func\-tion at the origin in configuration space.  Each element in (\ref{QQidentity}) is gauge invariant and renormalisation point independent.  The identity is exhibited in the heavy-quark limit of potential models for quarkonia; e.g., \cite{Lakhina:2006vg,olga}.  

In heavy-light systems it can be shown algebraically \cite{Ivanov:1998ms,Ivanov:1997yg,Ivanov:1997iu} that (\ref{QQidentity}) is realised via 
\begin{equation}
\label{hlmQE}
\rho_{\pi_{n}}^{\bar Q q}\propto (m_{\pi_{n}}^{\bar Qq})^{1/2} \; \& \; f_{\pi_{n}}^{\bar Q q}\propto 1/(m_{\pi_{n}}^{\bar Qq})^{1/2}.
\end{equation}
These results follow for heavy-light mesons because the integrands in (\ref{fpin}) \& (\ref{cpres}) can in this instance be accurately approximated via an expansion in $\varepsilon_{\pi_n}^{\bar Qq}/M^S_Q$ and $w_{\pi_n}^{\bar Q q}/M^S_Q$, where $w_{\pi_n}^{\bar Q q}$ is the width of the meson's Bethe-Salpeter wave function, which we define as the value of the relative momentum whereat the first Chebyshev moment of the amplitude ${\cal E}_{\pi_n}^{\bar Qq}(k;P)$ falls to one-half of its maximum value.  In the heavy-light meson $k\sim w_{\pi_n}^{\bar Q q}$ is the typical momentum of the light-quark.  Moreover, $w_{\pi_n}^{\bar Q q}$ obtains a finite nonzero value in the limit $M^S_Q \to \infty$.  It follows that a heavy-light meson is always of nonzero spatial extent.

\begin{figure}[t]
\centerline{\includegraphics[width=0.40\textwidth]{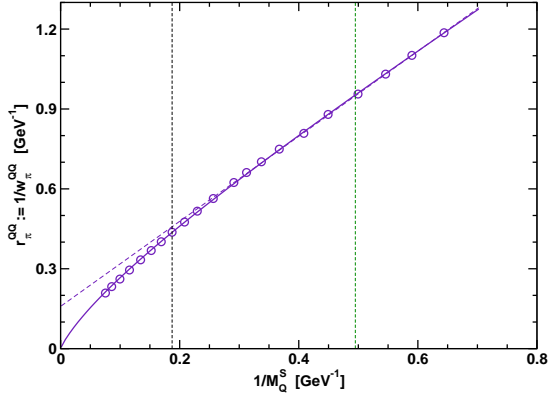}}

\caption{\label{fig:1} $r_{\pi_n}^{\bar QQ}$ vs.\ $1/M_{Q}^S$. Circles -- $r_{\pi_n}^{\bar QQ}$ calculated using the interaction model in \protect\cite{Maris:2002mt}.  Solid curve -- described in (\protect\ref{fig1curve}).  Dashed curve -- linear fit to calculated result in neighbourhood of $M_c^S$.  Dashed vertical lines mark, from left, $1/M_b^S$ \& $1/M_c^S$.  $r_{\pi_n}^{\bar QQ}$, 
approaches zero as \mbox{$M^Q_S \to \infty$}.  
}
\end{figure}

This is not true for heavy-heavy systems, as is apparent in Fig.\,\ref{fig:1}, which depicts the evolution of the spatial size of a heavy-heavy meson with constituent-quark spectrum-mass.\footnote{The evolution was calculated using the renormalisation-group-improved rainbow-ladder DSE truncation with the interaction model in \protect\cite{Maris:2002mt}.  All corrections to this truncation vanish in the heavy-heavy limit; e.g., see \protect\cite{Bhagwat:2004hn}.}  The curve in the figure is
\begin{equation}
\label{fig1curve}
 r_{\pi_n}^{{\bar QQ}} = \frac{\gamma_M}{M_Q^S}\,\ln \left[\tau_M+\frac{M_{Q}^E}{\Lambda_{\rm QCD}}\right]\,,\;
 \gamma_M = 0.68\,, \;\tau_M = 8.56\,,
\end{equation}
with $\Lambda_{\rm QCD}=0.234\,$GeV \cite{Maris:2002mt}.  Plainly, with increasing con\-sti\-tu\-ent-mass a heavy-heavy system becomes ``point-like'' in configuration space and hence delocalised in momentum space.  The evolution with $M_Q^S$ of an observable such as $f_{\pi_n}^{\bar QQ}$ may therefore be sensitive to the $\bar Q Q$ interaction over a wide range of momentum scales and hence a useful probe of that interaction.  

In NRQCD the matrix elements for various spin states of a given quarkonium system are equal up to corrections of order $v_Q^2 \simeq (w_{\pi_n}^{\bar QQ}/M^S_Q)^2$, where $k\sim w_{\pi_n}^{\bar QQ}$ is the typical magnitude of the heavy-constituent's three-momentum in the meson's rest frame \cite{Bodwin:1994jh}.  In this picture, $0^{-+}$ and $1^{--}$ mesons, which differ because spins are anti-aligned in the pseudoscalar and aligned in the vector, become degenerate in the limit $M^S_Q \to \infty$ and their leptonic decay constants become identical; i.e., $f_{\pi_n}^{\bar QQ}=f_{\rho_n}^{\bar QQ}$.  It is noteworthy, however, that (\ref{fig1curve}) gives $v_c^2\simeq 0.27$ and $v_b^2\simeq 0.18$.  Moreover, $v^2_Q$ falls only as $\alpha_s^2(M_{Q}^S)$.  Hence, a quantitative discrepancy between $f_{\pi_n}^{\bar QQ}$ \& $f_{\rho_n}^{\bar QQ}$ can conceivably persist until rather large quark masses.

The evolution to mass degeneracy is exemplified in \cite{Bhagwat:2004hn}, which begins with a model for the dressed-quark-gluon vertex that appears in (\ref{gensigma}).  Since the model's diagrammatic content is explicitly enumerable, a symmetry-preserving dres\-sed-quark Bethe-Salpeter kernel could be explicitly constructed.  The study showed that with rising cur\-rent-quark mass the rainbow-ladder truncation provides an increasingly accurate estimate of the mass of a heavy-heavy system, and the mass splitting between vector and pseudoscalar meson masses vanishes.  With the $b$-quark mass fitted to give $m_{\Upsilon(1S)}= 9.46\,$GeV, the model predicts $m_{\eta_b}= 9.42\,$GeV.  

\begin{figure}[t]
\centerline{\includegraphics[width=0.45\textwidth]{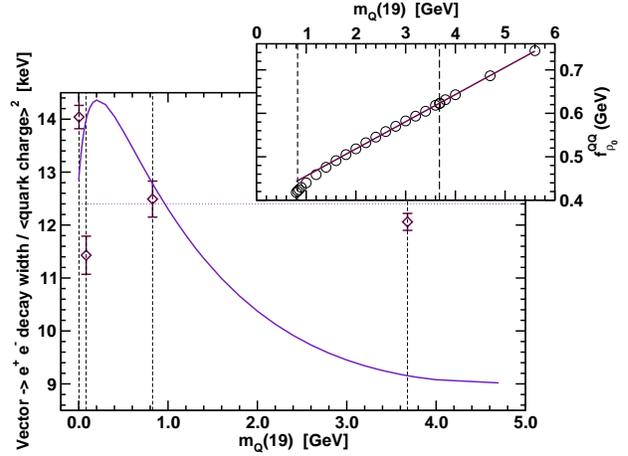}}

\caption{\label{fig:2} \textit{Main panel}. Width/charge-squared ratio in (\protect\ref{yennie}).  Circles -- result obtained in the approach of \protect\cite{Maris:2002mt}; diamonds -- data summarised in \protect\cite{Achasov:2001sn} and updated from \protect\cite{Yao:2006px}; horizontal dotted line indicates the conjecture of (\protect\ref{yennie}); and dotted vertical lines indicate, from left, the $u$, $s$, $c$, $b$ current-quark masses fixed at a renormalisation scale $\zeta=19\,$GeV.  \textit{Inserted panel}. Circles -- calculated evolution of $f_{\rho_0}^{\bar QQ}$.  Solid line -- straight line fit to large-$m_Q$ results.  The $c$ \& $b$ current-quark masses are indicated by vertical dashed lines.}
\end{figure}

Little is known experimentally about heavy-heavy $0^{-+}$ me\-sons.  However, $1^{--}$ mesons are readily produced experimentally and much studied.  We therefore consider vector meson leptonic decays, for which it was noted \cite{Yennie:1974ga,Achasov:2001sn} that  
\begin{equation}
\label{yennie}
 \frac{ ( f_{\rho_0}^{\bar QQ})^2}{m_{\rho_0}^{\bar QQ}} \propto \frac{\Gamma_{\rho_0}^{\bar QQ} \to e^+ e^-}{\langle e_Q\rangle^2} 
\approx 12.4\,{\rm keV},
\end{equation}
where $\langle e_Q\rangle$ is the mean electric charge of the valence-quark constituents in units of the electron's charge.  (See Fig.\,\ref{fig:2}.)  From this empirically based conjecture one might conclude that on the experimentally accessible domain of current-quark masses $f_{\rho_0}^{\bar QQ} \propto (m_{\rho_0}^{\bar QQ})^{1/2}$.  This is a marked departure from the behaviour in heavy-light systems, (\ref{hlmQE}), which is independent of the $\bar Q q$ interaction.  Furthermore \cite{Achasov:2001sn}, Coulomb-potential models typically give $f_{\rho_0}^{\bar QQ} \propto m_{\rho_0}^{\bar QQ}$ whereas a linear potential produces $f_{\rho_0}^{\bar QQ} \sim \,$constant.  Here is confirmation that the properties of quarkonia are a probe of the $\bar Q Q$ interaction.

In Fig.\,\ref{fig:2} we depict the result obtained for the ratio in (\protect\ref{yennie}) using the renormalisation-group-improved rainbow-ladder DSE truncation with the interaction model of \cite{Maris:2002mt}.  The interaction is precisely that of QCD in the ultraviolet; namely, colour-Coulomb.  However, a single parameter is employed to express a model for the long-range part of the quark-quark scattering kernel.  That parameter is a gluon mass-scale $m_g = 720\,$MeV.  It was chosen in order to fit $f_{\pi_0}$ \& $\rho_{\pi_0}$ and hence the results in Fig.\,\ref{fig:2} are an untuned prediction.  Quantitatively, that for the $c$-quark is good.  Indeed, for ground-state vector mesons up to and including $J/\psi$ the standard deviation between the calculated width and experiment is $\nlsim 15\,$\%.  (NB.\ The calculated masses are used in determining the widths.)  Adding the $\Upsilon(1S)$, that standard deviation rises to $\nlsim 25\,$\%.  Qualitatively, however, the mass-dependence obtained with this interaction does not support (\ref{yennie}).  Our preliminary result, illustrated via the inset in Fig.\,\ref{fig:2}, is $f_{\rho_0}^{\bar QQ} \propto m_{\rho_0}^{\bar QQ}$ for masses in the neighbourhood of the $b$-quark and beyond.  This is Coulomb-potential-like behaviour, which may be natural because the interaction employed is precisely that of QCD in the ultraviolet.  We are continuing to examine the validity of the hypothesis in (\ref{yennie}).

\section{Quark orbital angular momentum}
\label{qoam}
It is noteworthy that quark orbital angular momentum is not a Poincar\'e invariant.  However, if absent in a particular frame, it will inevitably appear in another frame related via a Poincar\'e transformation.  
Nonzero quark orbital angular momentum is thus a necessary outcome of a Poincar\'e covariant description, which is why the Bethe-Salpeter wave function in (\ref{genpibswf}) is a matrix-valued function with a rich structure.

A pseudoscalar meson naturally possesses total spin $J=0$ and this is expressed in the fact that $\chi_{\pi_n}$ is an eigenstate of the Pauli-Lubanski operator with eigenvalue zero.   Nonetheless, in the meson's rest frame one can straightforwardly decompose the Pauli-Lubanski operator into the sum of two terms: one measuring the angular momentum of the quarks and another measuring their spin.  In this way one can show that the terms in (\ref{genpibswf}) characterised by ${\cal E}$ and ${\cal F}$ are purely $L=0$ in the rest frame, whereas the ${\cal G}$ and ${\cal H}$ terms are associated with  $L=1$.  Thus a pseudoscalar meson Bethe-Salpeter wave function \emph{always} contains both $S$- and $P$-wave components.

In this connection, it is worth recalling that $E_{\pi_0}$ in (\protect\ref{genpibsa}) provides only a subleading contribution to the ultraviolet behaviour of the electromagnetic pion form factor \cite{Roberts:1994hh}.  The leading power-law behaviour anticipated from perturbative QCD is produced by the amplitudes associated with $F_{\pi_0}$ and $G_{\pi_0}$ \cite{Maris:1998hc}.  Moreover, it is an identity relating $F_{\pi_n}$ and $G_{\pi_n}$ in the ultraviolet that ensures $f_{\pi_n}$ in (\protect\ref{fpin}) is gauge invariant, and cutoff- and renormalisation-point-independent \cite{Maris:1998hc,Maris:1999nt}.

\begin{figure}[t]
\centerline{\includegraphics[width=0.45\textwidth]{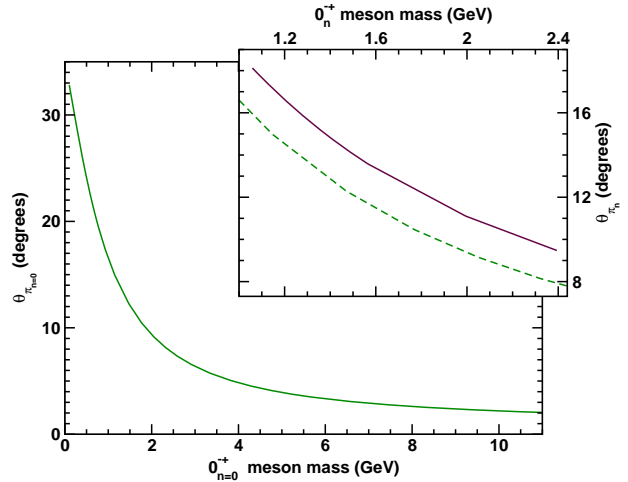}}

\caption{\label{fig:3} \emph{Main frame}: $\theta_{\pi_0}$, which gauges the rest-frame admixture of $L=1$ components in the ground-state $0^{-+}$ meson's Bethe-Salpeter wave function, plotted as a function of the meson's mass.  \emph{Inset}: Solid curve -- analogue, $\theta_{\pi_1}$, for the first excited state; dashed curve, $\theta_{\pi_0}$ for comparison.  In the chiral limit the ground state is naturally massless whereas the calculated mass of the first excited state is $1.04\,$GeV \protect\cite{Holl:2005vu}. }
\end{figure}

We exhibit the rest-frame angular momentum content of a $0^{-+}$ meson in the following way.  A Bethe-Salpeter amplitude is canonically normalised.  There are sixteen distinguishable terms in the associated sum; viz., an $\bar{\cal E}_{\pi_n} {\cal E}_{\pi_n}$ contribution plus an $\bar{\cal E}_{\pi_n} {\cal F}_{\pi_n}$ contribution, etc.  In the sum of the squares of these terms we associate $\sin^2\theta_{\pi_n}$ with the nondiagonal contributions, in which case $\sin\theta_{\pi_n}$ gauges the role played by $L=1$ components in the normalisation.  In Fig.\,\ref{fig:3}, for both the ground and first radially excited state, we plot the bound-state mass-dependence of $\theta_{\pi_n}$ obtained in the renormalisation-group-improved rainbow-ladder truncation using the model interaction described in \cite{Maris:2002mt}.  For both states, angular momentum is most significant in the neighbourhood of the chiral limit, and decreases with increasing current-quark mass.  Notably, as measured by the angle $\theta_{\pi_n}$, at a given bound-state mass the admixture of $L=1$ components in the first radial excitation is roughly 15\% greater than that in the ground state.  
Measured as a function of the current-quark mass, however, the situation is reversed: $\theta_1/\theta_0\approx 0.5$ in the chiral limit and this ratio increases steadily to $\approx 0.9$ at $m_Q(19)=0.5\,$GeV.  Our analysis continues.

\section{Baryons}
Despite material progress with the study of mesons, the challenge of baryons remains.  A nucleon appears as a pole in a six-point quark Green function.  The pole's residue is proportional to the nucleon's Faddeev amplitude, which is obtained from a Poincar\'e covariant Faddeev equation that adds-up all possible quantum field theoretical exchanges and interactions that can take place between three dressed-quarks.  In formulating and solving this problem, current expertise is approximately at the level it was for mesons ten years ago; i.e., model building and phenomenology.  However, we are a little ahead because a great deal has been learnt in applications to mesons.  For example, we have acquired a veracious understanding of the structure of dressed-quarks and -gluons and therefore can straightforwardly incorporate effects owing to and arising from the strong infrared modification of the momentum dependence of these propagators.  (See, e.g., Sect.\,5.1 of \cite{HUGS05}.)

A tractable treatment of the Faddeev equation requires a truncation.  One is based \cite{Cahill:1988dx} on the observation that an interaction which describes colour-singlet mesons also generates quark-quark (diquark) correlations in the colour-$\bar 3$ (antitriplet) channel \cite{Cahill:1987qr}.  The dominant correlations for ground state octet and decuplet baryons are $0^+$ and $1^+$ diquarks.  This can be understood on the grounds that: the associated mass-scales are smaller than the baryons' masses \cite{Burden:1996nh,Maris:2002yu}, with models giving (in GeV)
$m_{[ud]_{0^+}} = 0.74 - 0.82$,
$m_{(uu)_{1^+}}=m_{(ud)_{1^+}}=m_{(dd)_{1^+}}=0.95 - 1.02$;
the electromagnetic size of these correlations is less than that of the proton \cite{Maris:2004bp} -- $r_{[ud]_{0^+}} \approx 0.7\,{\rm fm}$, from which we estimate $r_{(ud)_{1^+}} \sim 0.8\,{\rm fm}$ based on the $\rho$-meson/$\pi$-meson radius-ratio \cite{Burden:1995ve,Hawes:1998bz}; and the positive parity of the correlations matches that of the baryons.  Both $0^+$ and $1^+$ diquarks provide attraction in the Faddeev equation.

The truncation of the Faddeev equation's kernel is completed by specifying that the quarks are dressed, with two of the three dressed-quarks correlated always as a colour-$\bar 3$ diquark.  Binding is then effected by the iterated exchange of roles between the bystander and diquark-participant quarks.  This ensures that the Faddeev amplitude exhibits the correct symmetry properties under fermion interchange.  A Ward-Takahashi-identity-pre\-ser\-ving electromagnetic current for the baryon thus constituted is subsequently derived~\cite{Oettel:1999gc}.  It depends on the electromagnetic properties of the axial-vector diquark correlation: its magnetic and quadrupole moments; and the strength of electromagnetically induced axial-vector $\leftrightarrow$ scalar diquark transitions.

A Faddeev equation study of the nucleon's mass and the effect on this of a pseudoscalar meson cloud are detailed in \cite{Hecht:2002ej}.  The lessons learnt are employed in a series of studies of nucleon properties; e.g., the nucleons' $\sigma$-term in \cite{Flambaum:2005kc,Holl:2005st}, and nucleon form factors in \cite{Alkofer:2004yf,Holl:2005zi,Bhagwat:2006py,Holl:2006zw}.  Of particular contemporary interest is the ratio $\mu_p G_E^p(Q^2)/G_M^p(Q^2)$ \cite{gao03,arrington}.  This passes through zero at $Q^2\approx 6.5\,$GeV$^2$ \cite{Holl:2005zi}.  The analogous ratio for the neutron is presented in \cite{Bhagwat:2006py}.  In the neighbourhood of $Q^2=0$, 
\begin{equation}
\label{smallQ}
\mu_p\,\frac{ G_E^n(Q^2)}{G_M^n(Q^2)} = - \frac{r_n^2}{6}\, Q^2 ,
\end{equation}
where $r_n$ is the neutron's electric radius.  The calculation shows this to be a good approximation for $r_n^2 Q^2 \nlsim 1$, with which the data \cite{Madey:2003av} are consistent.  It is notable that, just as for the proton, the small $Q^2$ behaviour of this ratio is materially affected by the neutron's pion cloud.  

Pseudoscalar mesons are not pointlike and therefore their contributions to form factors diminish in magnitude with increasing $Q^2$.  It follows therefore that the evolution of $\mu_p G_E^p(Q^2)/G_M^p(Q^2)$ and $\mu_n G_E^n(Q^2)/G_M^n(Q^2)$ on $Q^2\ngsim 2\,$GeV$^2$ are both primarily determined by the quark-core of the nucleon.  While the proton ratio decreases uniformly on this domain \cite{Alkofer:2004yf,Holl:2005zi}, \cite{Bhagwat:2006py} predicts that the neutron ratio increases steadily until $Q^2\simeq 8\,$GeV$^2$.  
 
As with mesons, Sect.\,\ref{qoam}, in a Poincar\'e covariant treatment the nucleon's quark core is necessarily described by a Faddeev amplitude with nonzero quark orbital angular momentum.  The Faddeev amplitude is therefore a matrix-valued function that, in a baryons' rest frame, corresponds to a relativistic wave function with $S$-wave, $P$-wave and $D$-wave components \cite{Oettel:1998bk}.  In form factor studies \cite{Alkofer:2004yf,Holl:2005zi,Bhagwat:2006py} there is some quantitative sensitivity to the electromagnetic structure of the diquarks.  However, the gross features of the form factors are primarily governed by correlations expressed in the nucleon's Faddeev amplitude and, in particular, by the amount of intrinsic quark orbital angular momentum \cite{Bloch:2003vn}.  The nature of the kernel in the Faddeev equation specifies just how much quark orbital angular momentum is present in a baryon's rest frame.  

We see a baryon as composed primarily of a quark core, constituted of confined quark and confined diquark correlations, but augmented by $0^-$ meson cloud contributions that are sensed by long wavelength probes.  Short wavelength probes pierce the cloud, and expose spin-isospin correlations and quark orbital angular momentum within the baryon.  The veracity of this description makes plain that a picture of baryons as a bag of three constituent-quarks in relative $S$-waves is profoundly misleading.

\section{Prospect}
Two emergent phenomena are primarily responsible for the observed properties of hadrons: confinement and dynamical chiral symmetry breaking (DCSB).  They can be viewed as an essential consequence of the presence and role of particle-antiparticle pairs in an asymptotically free theory and therefore can only be veraciously understood in relativistic quantum field theory.  The Dyson-Schwinger equations (DSEs) provide a natural framework for the exploration of these phenomena.  The DSEs are a generating tool for perturbation theory and thus give a clean connection with processes that are well understood.  Moreover, they admit a systematic, symmetry preserving and nonperturbative truncation, and thereby give access to strong QCD in the continuum.  On top of this, quantitative comparisons and feedback between DSE and lattice-QCD studies are today proving fruitful.

Dynamical chiral symmetry breaking (DCSB) is a singularly effective mass generating mechanism.  It is understood via QCD's gap equation, the solution of which delivers a quark mass function with a momentum-dependence that connects the perturbative and nonperturbative, con\-sti\-tuent-quark domains.  Despite the fact that light-quarks are made heavy, the mass of the pseudoscalar mesons remains unnaturally small.  That, too, owes to DCSB, expressed this time in a relationship between QCD's gap equation and those colour singlet Bethe-Salpeter equations which have a pseudoscalar projection.  Goldstone's theorem is a natural consequence of this connection.

The existence of a sensible truncation scheme enables the proof of exact results using the DSEs.  That the truncation scheme is also tractable provides a means by which the results may be illustrated, and furthermore a practical tool for the prediction of observables that are accessible at contemporary experimental facilities.  The consequent opportunities for rapid feedback between experiment and theory brings within reach an intuitive understanding of nonperturbative strong interaction phenomena.

There are indications that confinement may be expressed in the analyticity properties of the dressed propagators.  (Sect.\,4.2 in \cite{HUGS05}.)  To build understanding it is essential to work toward an accurate map of the confinement force between light-quarks and elucidate how that evolves from the potential between two static quarks.  Among the rewards are a clear connection between confinement and DCSB, an accounting of the distribution of mass within hadrons, and a realistic picture of hybrids and exotics.



\end{document}